\documentclass{article}
\usepackage[T1]{fontenc}
\usepackage[utf8]{inputenc}
\usepackage{booktabs}
\usepackage{multirow}
\usepackage{makecell}
\usepackage{ismir}
\usepackage{amsmath,cite,url,amssymb}
\usepackage{graphicx}
\usepackage{color}
\usepackage[table]{xcolor} 
\usepackage{pifont}
\usepackage{array}
\usepackage{adjustbox}

\definecolor{proxyrow}{HTML}{EEF7EE}
\definecolor{oursrow}{HTML}{EAF2FB}
\definecolor{baserow}{HTML}{F3F3F3}
\definecolor{maskcolor}{HTML}{E8F0FE}
\definecolor{arcolor}{HTML}{FFF3E0}
\definecolor{clcolor}{HTML}{E8F5E9}

\title{What Makes a Good Layer? Assessing the Layer-Wise Intrinsic Properties of Music Foundation Models}

\multauthor
  {Angelos-Nikolaos Kanatas$^{1}$ \hspace{1cm} Yuexuan Kong$^{2,3}$ \hspace{1cm} Pablo Alonso-Jiménez$^{1}$} 
  {{\bf Xavier Serra$^{1}$ \hspace{1cm} Dmitry Bogdanov$^{1}$}\\
  $^{1}$ Music Technology Group, Universitat Pompeu Fabra, Barcelona, Spain\\
  $^{2}$ Deezer Research, Paris, France\\
  $^{3}$ Nantes Université, École Centrale Nantes, CNRS, LS2N, UMR 6004, F-44000 Nantes, France\\
  {\tt\small angelosnikolaos.kanatas@upf.edu}
  }

\def\authorname{A.-N. Kanatas, Y. Kong, P. Alonso-Jiménez, X. Serra, and D. Bogdanov}

\usepackage[bookmarks=false,pdfauthor={\authorname},pdfsubject={\pdfsubject},hidelinks]{hyperref}

\begin{document}

\maketitle

\begin{abstract}

\noindent
Music foundation models are commonly used as frozen audio feature extractors, yet selecting which layer to extract from remains largely heuristic. Current practice defaults to fixed depths or multi-layer fusion, with limited understanding of why certain layers transfer better across downstream tasks or how representation quality varies with depth and pre-training paradigm.
We conduct a systematic layer-wise analysis of 12 music foundation models spanning three pre-training paradigms (masked modeling, autoregressive modeling, and contrastive learning), characterizing their hidden representations through intrinsic geometric and transformation-based properties.
Correlating label-free representation-quality metrics with layer-wise performance across 15 downstream tasks, we find that several metrics track layer quality for genre classification, emotion recognition, automatic tagging, and beat tracking, albeit with varying strength across tasks and pre-training paradigms. However, all metrics fail on tonal tasks such as key estimation and chord recognition, indicating that no single property serves as a general proxy for representation quality across music information retrieval tasks.
To address this gap, we introduce a pitch-transposition equivariance measure that captures properties missed by these standard metrics, providing a consistent indicator of tonal quality across model families.
Finally, we show that intrinsic metrics can serve as effective proxies for layer selection, matching or outperforming trainable multi-layer fusion methods, particularly in limited-data settings.

\end{abstract}

\section{Introduction}\label{sec:introduction}

Music foundation models trained with masked modeling~\cite{DBLP:conf/iclr/LiYZMCYXLRBGDLC24,DBLP:conf/icassp/WonHL24,DBLP:journals/corr/abs-2501-01108,DBLP:conf/mm/Alonso-JimenezR25}, contrastive learning~\cite{DBLP:conf/icassp/WuCZHBD23,DBLP:conf/ismir/HuangJLGLE22,DBLP:journals/corr/abs-2502-12511}, and autoregressive generation~\cite{DBLP:journals/corr/abs-2005-00341,DBLP:conf/nips/CopetKGRKSAD23,DBLP:journals/tmlr/0001TIZLFS0CRTU25,DBLP:journals/corr/abs-2503-08638} have emerged as effective general-purpose feature extractors for music audio, enabling transfer across diverse music information retrieval (MIR) tasks~\cite{DBLP:conf/nips/YuanMLZCYZLHTDW23,DBLP:journals/corr/abs-2408-14340}.
A common evaluation protocol freezes a pre-trained model, extracts representations from either a single layer---typically the final or a manually chosen intermediate layer---or a weighted combination of layers, and trains a lightweight probe for each downstream task~\cite{DBLP:conf/nips/YuanMLZCYZLHTDW23,DBLP:conf/interspeech/YangCCLLLLSCLHT21,DBLP:conf/nips/TurianSKRSSMTVM21}. 
Although this protocol enables controlled comparisons of frozen representations, layer selection remains consequential yet largely heuristic.
We therefore lack a principled understanding of what distinguishes layers that transfer well across diverse downstream tasks, what representation properties emerge across depth, and how they differ across pre-training paradigms.

Importantly, downstream probing primarily reveals whether task-relevant information can be decoded from a representation---for example, whether target labels are linearly separable---rather than how the latent space itself is organized~\cite{DBLP:journals/coling/Belinkov22,DBLP:conf/iclr/AlainB17,DBLP:conf/acl/PimentelVMZWC20}. It does not directly characterize how high-dimensional geometric structure evolves with depth or how representations respond to controlled transformations. Recent work argues that downstream performance captures only one aspect of representation quality and does not by itself provide a holistic characterization~\cite{DBLP:conf/ijcnn/PlachourasGFQBP25}.
Probe scores can also be misleading: control-task and information-theoretic probing studies show that high probe performance can partly reflect probe capacity and extraction effort rather than solely how readily a target property is encoded, motivating careful interpretation of probing results~\cite{DBLP:conf/emnlp/HewittL19,DBLP:conf/emnlp/VoitaT20}. 
Related evidence from speech shows that frozen-encoder rankings can vary with the downstream architecture, suggesting that probe-based benchmarks partly reflect readout design~\cite{DBLP:conf/interspeech/ZaiemKPER23}.
MIR benchmarks compound these concerns with well-documented validity threats: dataset confounds such as artist overlap can substantially distort classification results~\cite{DBLP:journals/jiis/Sturm13,DBLP:journals/tismir/Rodriguez-Algarra19,DBLP:journals/corr/abs-2301-01578}, while limited inter-rater agreement in perceptual annotation tasks can impose an upper bound on the performance differences benchmarks can meaningfully distinguish~\cite{flexer_problem_2016}. 
Accordingly, we operationalize a ``good'' layer for a task as one yielding high downstream performance under a fixed probe and evaluation protocol; this measures task-specific transfer utility, not universal representation quality.

Intrinsic metrics, computed from hidden representations without downstream supervision, provide a complementary characterization of representation quality.
%
In self-supervised vision, RankMe links embedding effective rank to downstream transfer~\cite{DBLP:conf/icml/GarridoBNL23}, while LiDAR refines rank-based assessment with an objective-aware criterion for joint-embedding architectures~\cite{DBLP:conf/iclr/Thilak0SDGNSL24}.
In speech and general audio, rank-based measures support model assessment, early performance prediction, and scaling-law analysis~\cite{DBLP:conf/icassp/AldenehTHTL25,DBLP:conf/interspeech/WhettenMPDE25,DBLP:journals/corr/abs-2510-10948}.
In parallel, geometric analyses show that representation structure changes systematically, often non-monotonically, across depth: intrinsic dimension, anisotropy, and curvature reveal intermediate objective-dependent phases associated with abstraction, compression, and transferability~\cite{DBLP:conf/nips/AnsuiniLMZ19,DBLP:conf/nips/ValerianiDCLAC23,DBLP:conf/eacl/RazzhigaevMGODK24,DBLP:conf/iclr/ChengDKMYLB25,DBLP:conf/icml/SkeanAZPNLS25}.
Contemporaneous studies in audio and speech characterize frozen self-supervised encoders layer-wise through compression, geometry, and invariance~\cite{sadok2026insidessl}, and apply participation ratio and isotropy to zero-label layer selection~\cite{batra2026where}.
These findings suggest that useful layers may be partly identifiable from representation geometry. 
Concurrent work combines such geometry with a small labeled calibration set to select and fuse task-discriminative layers~\cite{batra2026latent}. 
Whether this extends to MIR remains unclear, since downstream music tasks impose qualitatively different representational demands.

Layer-wise analysis in music is still emerging.
Early work on Jukebox showed that codified-audio language-model representations transfer well to discriminative MIR tasks, with average performance peaking at intermediate layers~\cite{DBLP:conf/ismir/CastellonDL21}; synthetic probing found that music-theory decodability varies across concepts, layers, and model sizes in Jukebox and MusicGen~\cite{DBLP:conf/ismir/WeiFD024}.
For masked-prediction encoders, MusicFM and MuQ exhibit an acoustic-to-semantic depth trend, with task-appropriate single layers often outperforming trainable weighted all-layer fusion~\cite{DBLP:journals/corr/abs-2505-16306}.
Probing and intervention studies on MERT and MusicGen indicate that content-related attributes, such as pitch and chord root, become more consistently recoverable with depth than style-related attributes, such as timbre and chord quality~\cite{ma_music_symbolic}, while linear-probe gains from root- and third-note interventions in MusicGen suggest chord quality is encoded but not readily accessible~\cite{ma_exploring_2024}.
Head-wise probing, intermediate decoding, and interchange interventions trace layer- and component-level signatures of musical attributes in MusicGen~\cite{Koo2024apr,DBLP:conf/ismir/VasquezPBZ24}, while sparse autoencoders enable unsupervised concept discovery and reveal depth- and scale-dependent interpretability~\cite{singh_discovering_2026}.

Taken together, prior work relies primarily on task-specific probes, targeted interventions, or post-hoc feature discovery; our work instead characterizes how representation geometry and transformation behavior evolve across depth, and tests whether these properties can serve as label-free proxies for downstream layer quality.
To our knowledge, no prior study has systematically examined music-audio representations across layers and pre-training paradigms through their \textit{intrinsic properties}, nor evaluated whether representation-quality metrics adopted in other domains can also guide practical layer selection as an alternative to exhaustive layer scanning and trainable layer fusion across diverse MIR tasks.

We summarize our main contributions as follows:
\begin{enumerate}

    \item We present the first systematic layer-wise analysis of \textbf{intrinsic representation properties} in music foundation models, covering 12 models across masked, autoregressive, and contrastive pre-training paradigms, and identify distinct depth-wise representational profiles.

    \item We find that intrinsic metrics are task- and paradigm-dependent \textbf{proxies for downstream layer quality}: intrinsic dimension, curvature, and invariance-based metrics track layer-wise downstream performance across several tasks, but fail on tonality, motivating a \textbf{pitch-transposition equivariance} measure.

    \item We demonstrate that intrinsic metrics enable \textbf{label-free layer selection and fusion}, reducing exhaustive probing to a few candidate layers while recovering near-oracle downstream performance and matching or outperforming trainable multi-layer fusion, particularly when labeled data is scarce.
    
\end{enumerate}

\noindent Code and extended layer-wise results with per-model and per-task breakdowns are available on the project page.\footnote{\url{https://angeloskanatas.github.io/music-fms-layer-eval/}}


\section{Methodology}\label{sec:method}


\subsection{Models}\label{subsec:models}

We analyze 12 publicly available music foundation models spanning three pre-training paradigms (Table~\ref{tab:model_details}).
The \emph{masked} family comprises two MERT variants together with MusicFM, MuQ, and OMAR-RQ. These models differ primarily in how their masked prediction targets are constructed: MERT~\cite{DBLP:conf/iclr/LiYZMCYXLRBGDLC24} combines acoustic (EnCodec) and musical (CQT) teachers; MusicFM~\cite{DBLP:conf/icassp/WonHL24} follows \mbox{BEST-RQ} by quantizing log-mel spectra using a frozen random projection; MuQ~\cite{DBLP:journals/corr/abs-2501-01108} predicts Mel-RVQ tokens, with the MuQ$_{\text{iter}}$ variant employing a tokenizer retrained on latent representations from a first-pass MuQ model; and OMAR-RQ~\cite{DBLP:conf/mm/Alonso-JimenezR25} extends BEST-RQ to multi-feature masked token prediction, using targets derived from EnCodec, mel, CQT, and waveform representations.
The \emph{autoregressive} group covers five model scales across two decoder designs: MusicGen-S/M/L~\cite{DBLP:conf/nips/CopetKGRKSAD23} are EnCodec-token transformer decoders trained with text conditioning, and \mbox{YuE-0.5B/7B}~\cite{DBLP:journals/corr/abs-2503-08638} are LLaMA\,2 decoders trained for lyrics-to-song generation using X-Codec audio tokens together with textual control signals. For both autoregressive families, we extract hidden representations without external conditioning, feeding only audio token sequences.
The \emph{contrastive} models are trained with InfoNCE but differ along the cross-modal versus within-modal axis: \mbox{CLAP}~\cite{DBLP:conf/icassp/WuCZHBD23} aligns audio and natural-language captions in a shared embedding space, whereas Myna~\cite{DBLP:journals/corr/abs-2502-12511} treats paired masked views of the same track as positives, with token masking replacing traditional audio augmentations.

\begin{table}[t]
\centering
\small
\setlength{\tabcolsep}{3.5pt}
\renewcommand{\arraystretch}{1.12}
\begin{tabular}{@{}l @{\hskip 8pt} l @{\hskip 9pt} c @{\hskip 6pt} c @{\hskip 0.2pt} r@{}}
\toprule
\textbf{Model} & \textbf{Architecture} & $\boldsymbol{L}$ & $\boldsymbol{H}$ & \textbf{Params} \\
\midrule

\rowcolor{maskcolor}
\multicolumn{5}{@{}l}{\textit{\textbf{Masked Modeling (MM)}}} \\
MERT-v1-95M                 & Transformer Enc. & 12 & 768       & 95\,M \\
MERT-v1-330M                & Transformer Enc. & 24 & 1024      & 330\,M \\
MusicFM$_{\text{MSD}}$      & Conformer Enc.   & 12 & 1024      & 330\,M \\
MuQ$_{\text{iter}}$         & Conformer Enc.   & 12 & 1024      & 310\,M \\
OMAR-RQ$_{\text{multifeature}}$          & Conformer Enc.   & 24 & 1024      & 580\,M \\
\addlinespace[2pt]

\rowcolor{arcolor}
\multicolumn{5}{@{}l}{\textit{\textbf{Autoregressive (AR)}}} \\
MusicGen-S  & Transformer Dec. & 24 & 1024      & 300\,M \\
MusicGen-M  & Transformer Dec. & 48 & 1536      & 1.5\,B \\
MusicGen-L  & Transformer Dec. & 48 & 2048      & 3.3\,B \\
YuE-s1-0.5B & LLaMA\,2         & 24 & 1024      & 0.5\,B \\
YuE-s1-7B   & LLaMA\,2         & 32 & 4096      & 7\,B \\
\addlinespace[2pt]

\rowcolor{clcolor}
\multicolumn{5}{@{}l}{\textit{\textbf{Contrastive Learning (CL)}}} \\
LAION-CLAP & HTSAT\textsuperscript{\dag}    & 21 & 128--1024 & 73\,M \\
Myna-Base  & ViT-S/16 & 12 & 384       & 22\,M \\
\bottomrule
\end{tabular}
\caption{Models included in our study, grouped by pre-training paradigm. $L$: number of layers; $H$: hidden dim. \textsuperscript{\dag}\,Hierarchical Swin Transformer; $H$ doubles across stages.}
\label{tab:model_details}
\end{table}

\subsection{Intrinsic Metrics}\label{subsec:metrics_definition}

To compute the intrinsic metrics, we extract layer-wise representations from each frozen model using $N\!=\!10{,}000$ randomly sampled 15-second clips (one per track) from the MTG-Jamendo dataset~\cite{bogdanov2019mtg} (about 55,000 tracks).\footnote{In robustness checks, depth-wise metric profiles were stable over disjoint subsamples and appeared to converge by this sample size; across alternative corpora, absolute magnitudes can shift with corpus characteristics, whereas relative layer-wise patterns persist.} For each layer $l \in \{0,\dots,L\}$, we obtain a sequence of hidden states.
%
For sequence-level metrics, we aggregate each clip into a single vector: we mean-pool frame outputs over time for encoder-based models and use the final-token hidden state for autoregressive decoders because, under causal
attention, it attends to the full input sequence.
Stacking these embeddings across clips yields a representation matrix $\mathbf{Z}^{(l)} \in \mathbb{R}^{N \times H}$.
For frame-level metrics, we instead retain the full token sequence $\{\mathbf{z}_t^{(l)}\}$.

We characterize layers from three perspectives: (i) \emph{geometric and spectral} properties of the representation manifold, (ii) \emph{augmentation-conditioned} discriminability and invariance, and (iii) \emph{pitch-transposition equivariance}.

\vspace{2pt}

\noindent\textbf{Geometric and Spectral.}
The \emph{intrinsic dimension (ID)} of a representation is the minimum number of variables needed to describe its underlying data manifold.
We estimate ID using the TwoNN estimator~\cite{DBLP:journals/corr/abs-1803-06992}, which models the ratio $\mu_i = r_{i,2}/r_{i,1}$ of second- to first-nearest-neighbor distances under a local uniformity assumption. Layer-wise profiles were qualitatively consistent with those of GRIDE~\cite{denti2022gride}, which extends TwoNN to higher-order neighbor ratios and scale-dependent analysis, and PHD~\cite{DBLP:journals/dcg/Schweinhart21}, a topological estimator based on persistent homology. 
While ID captures the degrees of freedom of the manifold, it does not reveal how variance is distributed across those dimensions.
\emph{Effective rank (RankMe)}~\cite{DBLP:conf/eusipco/RoyV07,DBLP:conf/icml/GarridoBNL23} exponentiates the Shannon entropy of the singular values of $\mathbf{Z}^{(l)}$, normalized as $p_k = \sigma_k / \|\boldsymbol{\sigma}\|_1$, yielding a continuous measure of how many dimensions are effectively used: lower values indicate a spectrum concentrated in few directions, whereas higher values reflect a more uniformly distributed spectrum.
%
%
\emph{Anisotropy}~\cite{DBLP:conf/eacl/RazzhigaevMGODK24} measures the fraction of total variance concentrated along the dominant singular direction ($\sigma_1^2 / \sum_i \sigma_i^2$), capturing the degree to which a representation collapses toward a low-dimensional subspace.
Finally, temporal \emph{curvature}, introduced in studies of perceptual straightening in visual neuroscience~\cite{henaff2019perceptual} and later adopted as a layer-wise diagnostic for token-sequence trajectories in language models~\cite{DBLP:conf/nips/HosseiniF23,DBLP:conf/icml/SkeanAZPNLS25,DBLP:journals/corr/abs-2601-22364}, quantifies how sharply a representation trajectory bends over time.
For each clip and layer, we form frame-level displacement vectors $\mathbf{v}_t^{(l)} = \mathbf{z}_{t+1}^{(l)} - \mathbf{z}_t^{(l)}$ and measure the mean angle between successive displacements,
\begin{equation*}
C^{(l)} \;=\; \mathbb{E}_{\mathbf{x},\,t}\!\left[\,\arccos\!\big(\hat{\mathbf{v}}_{t+1}^{(l)} \cdot \hat{\mathbf{v}}_{t}^{(l)}\big)\,\right] \;\in\; [0,\pi],
\end{equation*}
with $\hat{\mathbf{v}} = \mathbf{v}/\lVert\mathbf{v}\rVert$ and the expectation taken over clips $\mathbf{x}$ and successive displacement pairs within each clip.

\vspace{2pt}

\noindent\textbf{Augmentation-Conditioned.}
The metrics above characterize static layer geometry but not how representations respond to controlled input perturbations.
To assess this behavior, we obtain augmented views of each clip by applying pitch shifting ($\pm 4$~semitones), time stretching ($\times 0.85$--$1.15$), additive Gaussian noise, gain variation, time shift, and low-pass filtering.
For each downstream task family, we exclude augmentations that alter task-defining attributes (see Section~\ref{subsec:downstream}): pitch shifts for tonal tasks and time stretching for rhythm tasks.
\mbox{\emph{LiDAR}} (Linear Discriminant Analysis Rank)~\cite{DBLP:conf/iclr/Thilak0SDGNSL24} 
adapts linear discriminant analysis (LDA) to joint-embedding self-supervised learning by treating each original clip as a surrogate class and its augmented views as within-class samples; we use 10 views per clip. It forms between- and within-clip scatter matrices, $\boldsymbol{\Sigma}_b$ and $\boldsymbol{\Sigma}_w$, and computes the effective rank of the LDA matrix $\boldsymbol{\Sigma}_w^{-1/2}\boldsymbol{\Sigma}_b\boldsymbol{\Sigma}_w^{-1/2}$. 
The resulting score estimates how many directions discriminate clips after whitening augmentation-induced variability, discounting high-variance directions unstable under the chosen augmentations.
\emph{InfoNCE}~\cite{DBLP:journals/corr/abs-1807-03748} complements LiDAR by measuring agreement between augmented views relative to negatives: for each clip, two augmented views form a positive pair scored against other clips in the batch, with lower loss indicating stronger invariance to the chosen augmentations.

\begin{figure*}[t]
  \centering
  \begin{minipage}[c]{0.665\textwidth}
    \centering
    \includegraphics[width=\textwidth]{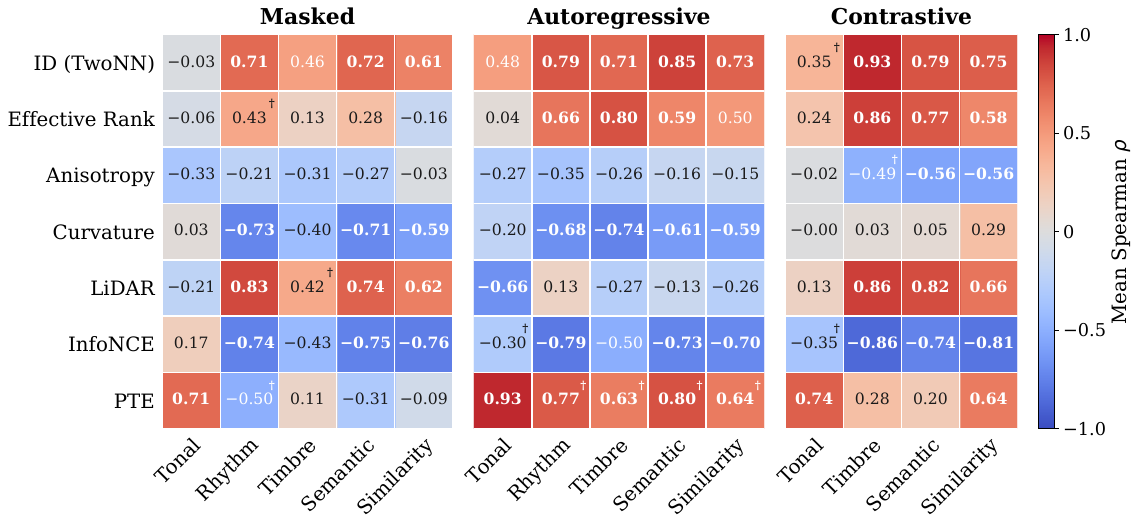}
    \centerline{\small (a) Metric--task correlations by paradigm}
  \end{minipage}%
  \hspace{0.01\textwidth}%
  \begin{minipage}[c]{0.325\textwidth}
    \centering
    \vspace{12pt}
    \includegraphics[width=\textwidth]{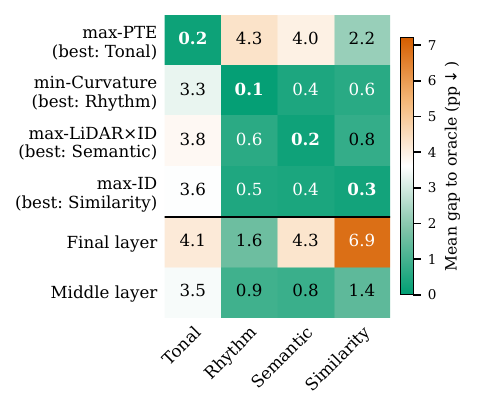}
    \par\vspace{1.2pt}
    \centerline{\small (b) Proxy-guided layer selection}
  \end{minipage}
  \caption{(a) Mean within-model Spearman~$\rho$ between intrinsic-metric and downstream-performance layer profiles, averaged by pre-training paradigm. $^\dagger$\,Weakens or flips sign under depth control (partial Spearman) or across models. (b) Top-3 proxy-guided layer-selection mean gap (pp$\downarrow$) for masked models. The best intrinsic-metric proxy varies by task.}
  \label{fig:metrics-tasks}
\end{figure*}

\vspace{2pt}

\noindent\textbf{Pitch-Transposition Equivariance (PTE).}
Tonal tasks (key estimation, chord recognition, and pitch classification) rely on transposition-equivariant structure, which the metrics above do not assess.
We therefore introduce \emph{PTE}, a self-supervised diagnostic adapted from the cross-power spectral density objective of STONE~\cite{DBLP:conf/ismir/KongLMWLH24,DBLP:journals/spl/LostanlenKMLH25}.
At each layer $l$, a linear probe $h_\phi$ maps the sequence-level embedding to a 12-dimensional softmax key signature profile, projected onto the circle of fifths by the DFT at $\omega\!=\!7$.
The same probe is applied independently to a clip $\mathbf{x}$ and its audio-domain $k$-semitone transposition $T_k(\mathbf{x})$; for an equivariant probe, the cross-power phase equals $-2\pi\omega k/12$.
We train $h_\phi$ against this target on 10,000 MTG-Jamendo clips (70/15/15 split by clip before shifting) using 11 nonzero shifts, with Adam (learning rate $10^{-3}$) for up to 200 epochs with early stopping.
Each layer is scored by the RMS chordal distance $d^{(l)}$ between the unit-normalized cross-power and its target, reported as $\mathrm{PTE}^{(l)} = 1 - \tfrac{1}{2}\,d^{(l)} \in [0,1]$, where a high score indicates 12-TET transposition structure is decodable from the frozen representation.

\subsection{Probing and Downstream Tasks}\label{subsec:downstream}

We evaluate layer-wise representations across a diverse set of downstream tasks organized into five families spanning complementary aspects of music: pitch and tonality, rhythm, timbre, high-level semantics, and perceptual similarity.
\emph{Tonal} comprises key estimation on \mbox{GiantSteps-Key}~\cite{DBLP:conf/ismir/KneesFHVBHG15} (\texttt{mir\_eval}~\cite{DBLP:conf/ismir/RaffelMHSNLE14} weighted score), pitch classification on NSynth~\cite{DBLP:conf/icml/EngelRRDNES17} (top-1 accuracy), and chord recognition on a 1,217-song corpus from the Isophonics, Billboard, and MARL collections~\cite{DBLP:conf/ismir/JiangCLX19} using a decomposed probe with independent root, bass, and pitch-class activation heads~\cite{mcfee2017structured,DBLP:conf/ismir/PoltronieriSR25} (\texttt{mir\_eval} root, triads, and MIREX).
\emph{Rhythm} includes beat and downbeat tracking on the GTZAN-Rhythm annotations~\cite{marchand2015gtzanrhythm} (\texttt{mir\_eval} F-measure at the standard $70\,\mathrm{ms}$ tolerance).
\emph{Timbre} consists of NSynth instrument-family classification (top-1 accuracy).
\emph{Semantic} includes genre classification on GTZAN~\cite{DBLP:journals/taslp/TzanetakisC02} using the fault-filtered split~\cite{kereliuk2015adversaries} (top-1 accuracy); automatic tagging on \mbox{MagnaTagATune}~\cite{DBLP:conf/ismir/LawWMBD09} and four MTG-Jamendo~\cite{bogdanov2019mtg} subsets (top-50, genre, instrument, mood/theme), each scored by macro-averaged AP and ROC-AUC; emotion regression on EmoMusic~\cite{DBLP:conf/mm/SoleymaniCSSY13} ($R^2$ on arousal and valence); and clip-level structural-function recognition on the Harmonix Set~\cite{DBLP:conf/ismir/NietoMDRSE19} (top-1 accuracy).
\emph{Similarity} evaluates perceptual music similarity on Dim-Sim~\cite{DBLP:conf/icassp/LeeBSJN20}, a dataset of human-rated similarity triplets, using triplet agreement computed from pairwise distances between pooled representations, without training a probe.
For all supervised tasks, the pre-trained model is frozen, and each layer's representations serve as input to a task-specific shallow MLP head with one 512-unit hidden layer, following the MARBLE constrained-track settings~\cite{DBLP:conf/nips/YuanMLZCYZLHTDW23}.
%


\section{Empirical Findings}\label{sec:results_discussion}

We first analyze how intrinsic representation properties relate to downstream performance across depth, tasks, and pre-training paradigms, and then evaluate their utility as proxies for layer selection and fusion.

\subsection{Correlation with Downstream Performance}\label{subsec:downstream_correlation}

\noindent\textbf{The distribution of task-relevant information varies across depth and pre-training paradigms.}
For the masked and autoregressive models studied, intermediate layers consistently outperform the final layer, in line with cross-domain observations in speech~\cite{DBLP:conf/icassp/PasadSL23}, text~\cite{DBLP:conf/icml/SkeanAZPNLS25}, and vision~\cite{DBLP:journals/corr/abs-2504-13181, DBLP:journals/corr/abs-2601-09322}.
Within masked encoders, downstream performance peaks follow a clear task-family depth ordering: tonal earliest (${\sim}30\%$), semantic mid-depth (${\sim}53\%$), and rhythm later (${\sim}64\%$), extending a prior two-model analysis~\cite{DBLP:journals/corr/abs-2505-16306} and paralleling layer-wise feature hierarchies in text~\cite{DBLP:conf/acl/TenneyDP19,DBLP:conf/acl/JawaharSS19} and speech~\cite{DBLP:journals/taslp/YangCHLLWSCTHFCLCHTLLMWL24,DBLP:conf/icassp/PasadSL23}.
In autoregressive models, most non-tonal tasks cluster within an intermediate-depth band that tends to tighten with scale, whereas tonal tasks shift toward the deepest layers.
In our contrastive encoders, non-tonal performance instead concentrates near the final layers, while tonal peaks earlier, consistent with the shallow-harmonic/deep-rhythmic organization reported in a contrastive music transformer~\cite{DBLP:conf/ismir/KongMLLH25}.
Yet these aggregate patterns capture only coarse, fragile family-level orderings and do not identify the best layer per model--task pair, even within a paradigm, motivating the intrinsic-metric analysis and layer-selection strategies below.

\vspace{2pt}

\begin{figure}[t]
  \centering
  \includegraphics[width=\columnwidth]{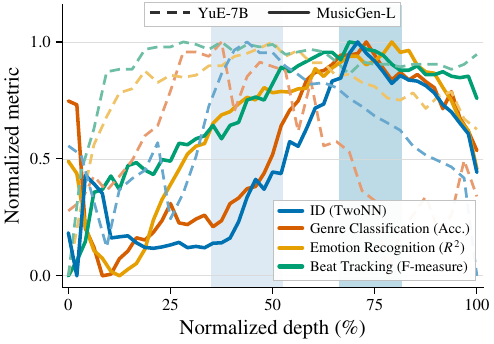}
  \caption{In autoregressive models, intrinsic dimension peaks within the shaded intermediate-depth band where performance on semantic and rhythm tasks is highest.}
  \label{fig:ar_id-downstream}
\end{figure}

\noindent\textbf{Intrinsic metrics predict non-tonal layer quality, but their utility varies by task and paradigm.}
To quantify metric--task alignment, we compute, for each model, the Spearman~$\rho$ between intrinsic-metric and downstream-performance layer profiles (Figure~\ref{fig:metrics-tasks}a).
%
ID emerges as the most consistent proxy for non-tonal layer quality, achieving the highest mean correlation overall ($\bar\rho{=}0.76$), with a consistent correlation sign across non-tonal task families in all pre-training paradigms.
%
Curvature is negatively correlated with downstream performance across rhythm, semantic, timbre, and similarity tasks in masked and autoregressive models, most consistently for beat and downbeat tracking (masked: $\bar\rho{=}{-}0.69/{-}0.76$; autoregressive: $\bar\rho{=}{-}0.68/{-}0.67$; negative in every model).
This indicates that layers with \emph{straighter} (lower-curvature) temporal trajectories tend to transfer better, broadly consistent with prior work linking representational straightening to improved prediction in autoregressive language models~\cite{DBLP:conf/nips/HosseiniF23}, better-performing layers for downstream text-embedding tasks~\cite{DBLP:conf/icml/SkeanAZPNLS25}, and gains from explicit straightening objectives in vision~\cite{DBLP:conf/nips/NiuSS24} and planning~\cite{DBLP:journals/corr/abs-2603-12231}.
These depth-wise patterns differ by paradigm.
In masked encoders, ID follows a broader inverted-$U$ profile and complements curvature: curvature aligns more strongly with rhythm, whereas ID is slightly more consistent across semantic tasks.
In autoregressive models, this dissociation largely disappears: although curvature remains predictive, ID follows a sharper inverted-$U$ profile that tightens with scale and aligns with the increasingly narrow task-relevant depth band.
ID accordingly becomes the strongest non-tonal proxy, especially at larger scales ($\rho{=}{+}0.65$ to ${+}0.97$; Figure~\ref{fig:ar_id-downstream}), consistent with recent language-model evidence that an intermediate high-ID phase marks the onset of downstream-transferable representations~\cite{DBLP:conf/iclr/ChengDKMYLB25}.
\mbox{LiDAR} and InfoNCE largely track ID and curvature in masked encoders (pairwise $|\rho|{=}0.61$--$0.90$), and the elementwise LiDAR${\times}$ID product yields the strongest semantic proxy ($\bar\rho{=}0.78$). In autoregressive models, \mbox{LiDAR} is architecture-dependent: it correlates positively in YuE but reverses sign in \mbox{MusicGen}.
Compared with ID and curvature, effective rank is weaker and paradigm-dependent: stronger in autoregressive and contrastive than in masked models, consistent with prior work on rank as a useful but limited within-model signal~\cite{DBLP:conf/icassp/AldenehTHTL25}. 
Anisotropy is sign-inconsistent and better viewed as an architecture-dependent property than a general quality signal~\cite{DBLP:conf/eacl/GodeyCS24,DBLP:conf/acl/SaadaN23}.
%

\vspace{2pt}

\noindent\textbf{PTE tracks tonal-task performance where standard metrics fail.}
Across models, standard intrinsic metrics show weak, sign-inconsistent within-model correlations with tonal-task performance, whereas PTE is the only consistently positive proxy, with the strongest overall alignment ($\bar\rho{=}0.80$) and per-task correlations: pitch ($\bar\rho{=}0.87$), key ($\bar\rho{=}0.83$), and chord recognition ($\bar\rho{=}0.59/0.63$ for root/triads).
We hypothesize that standard metrics fail because they characterize static geometry or augmentation-induced invariance rather than directly assessing whether transposition-equivariant structure is accessible.

\vspace{2pt}

\noindent\textbf{Metrics do not yield reliable cross-model rankings.}
The within-model analyses above show that intrinsic metrics can identify strong layers; we next ask whether raw metric values predict downstream model rankings. For each task, we correlate metric values with downstream performance across models at their respective best-performing layers; $\rho_{\mathrm{c}}$ denotes a cross-model Spearman correlation.
We find that effective rank is strongly correlated with hidden dimension ($\rho_{\mathrm{c}}\!=\!0.82$) but nearly unrelated to best-layer downstream quality across all models ($\rho_{\mathrm{c}}\!=\!0.02$), suggesting that raw values are not directly comparable across heterogeneous architectures.
%
%
Within masked encoders, effective rank and ID both track sequence-level downstream performance ($\rho_{\mathrm{c}}\!=\!0.70$), consistent with the within-family predictive power of effective rank reported for audio self-supervised models~\cite{DBLP:journals/corr/abs-2510-10948}. Yet this association is not stable: it varies across tasks and reverses sign for autoregressive models ($\rho_{\mathrm{c}}\!=\!-0.47$), reinforcing the task- and paradigm-dependence observed in our within-model analyses.
These limitations are consistent with known caveats: spectral metrics have mainly been proposed and validated within specific training paradigms~\cite{DBLP:conf/icml/GarridoBNL23,DBLP:journals/corr/abs-2510-10948,DBLP:conf/nips/AgrawalMGR22} and can be setting-dependent~\cite{DBLP:conf/tagml/TsitsulinMP23}; augmentation-conditioned metrics can entangle representation quality with model-specific learned invariances~\cite{DBLP:conf/iclr/Thilak0SDGNSL24}; and geometric metrics such as ID and curvature can be sensitive to the input distribution~\cite{DBLP:conf/iclr/ChengDKMYLB25,DBLP:journals/corr/abs-2601-22364}, requiring careful interpretation in cross-model comparisons.
Notably, PTE is the only metric in our study that retains a consistent tonal signal both within models and across models ($\rho_{\mathrm{c}}$: pitch $1.00$, key $0.75$, chord $0.80$).

\subsection{Proxy-Guided Layer Selection and Fusion}\label{subsec:layer_selection}

The within-model correlation analysis above motivates a downstream-supervision-free layer-selection strategy: for each task family, we retrospectively use the best-aligned metric as a proxy, rank layers accordingly, and evaluate only the top-ranked ones (Figure~\ref{fig:metrics-tasks}b). This reduces the search space to a small candidate set, offering a lightweight alternative to exhaustive layer-wise scanning.
We compare two proxy-based strategies against depth heuristics and both non-trainable and trainable all-layer fusion baselines (Table~\ref{tab:ls_comparison}): (i)~\emph{proxy selection}, which evaluates only the top-$k$ proxy-ranked layers by training one probe per layer ($k\!\in\!\{1,3\}$); and (ii)~\emph{proxy-guided fusion}, which combines those top-$k$ layers using a fixed operator (mean or concatenation) before training a single probe.
Among the trainable baselines, \emph{weighted sum} learns softmax-normalized scalar layer weights~\cite{DBLP:conf/naacl/PetersNIGCLZ18,DBLP:conf/interspeech/YangCCLLLLSCLHT21}, \emph{HConv} hierarchically aggregates layers with stacked 1D convolutions~\cite{DBLP:conf/interspeech/ShihH24}, and \emph{attentive fusion} applies cross-attention with a shared learnable query over per-layer representations~\cite{DBLP:journals/corr/abs-2601-09322}.
We report each method's mean gap in percentage points (pp) to the per-task \emph{oracle}---the best-performing single layer for each (model, task) pair under that task's evaluation protocol.

\begin{table}[t]
  \centering
  \footnotesize
  \setlength{\tabcolsep}{1.25pt}
  \renewcommand{\arraystretch}{1.08}
  \begin{adjustbox}{max width=\columnwidth}
  \begin{tabular}{@{}l >{\hspace{4pt}}c c >{\hspace{8.5pt}}c >{\hspace{8.5pt}}c >{\hspace{4.5pt}}c >{\hspace{3.5pt}}c@{}}
  \toprule
  & \multicolumn{2}{c}{\shortstack[c]{\textbf{Pre-Training}\\[-1pt]\textbf{Paradigm}}}
  & \multicolumn{2}{c}{\shortstack[c]{\textbf{Probe}\\[-1pt]\textbf{Train Size}}}
  & & \\
  \cmidrule(lr{6pt}){2-3}\cmidrule(lr{6pt}){4-5}
  \textbf{Method}
  & \multicolumn{1}{c}{\hspace{1.5pt}\textbf{Masked}}
  & \textbf{\hspace{1.5pt}AR}
  & \textbf{\hspace{-7pt}$<\!1$K}
  & \textbf{\hspace{-7pt}$\geq\!6$K}
  & \textbf{Overall}
  & \textbf{\% $\geq$ Orac.} \\
  \midrule
  \addlinespace[2pt]
  \multicolumn{7}{@{}l}{\textbf{Proxy-Guided Layer Selection}} \\[2pt]
  \rowcolor{proxyrow}
  Top-1 candidate
    & 0.6 & 0.9 & 1.8 & 0.7 & 1.0 & 15 \\
  \rowcolor{proxyrow}
  Top-3 candidates
    & \textbf{0.2} & \underline{0.4} & \textbf{0.4} & \underline{0.2} & \underline{0.4} & 58 \\
  \addlinespace[5pt]
  \multicolumn{7}{@{}l}{\textbf{Non-Trainable Multi-Layer Fusion}} \\[2pt]
  \rowcolor{oursrow}
  Proxy-guided avg. (top-3)
    & \underline{0.4} & \textbf{0.1} & \underline{1.3} & \textbf{0.1} & \textbf{0.3} & 41 \\
  \rowcolor{oursrow}
  Proxy-guided concat. (top-3)
    & 0.5 & 1.1 & 2.1 & 0.4 & 0.8 & 31 \\
  All-layer avg.
    & 1.2 & 3.2 & 5.4 & 0.8 & 2.0 & 24 \\
  All-layer concat.
    & 2.6 & 4.1 & 5.8 & 2.7 & 3.6 & 9 \\
  \addlinespace[5pt]
  \multicolumn{7}{@{}l}{\textbf{Trainable Multi-Layer Fusion}} \\[2pt]
  Weighted sum~\cite{DBLP:conf/naacl/PetersNIGCLZ18,DBLP:conf/interspeech/YangCCLLLLSCLHT21}
    & 0.9 & 2.8 & 5.6 & 0.5 & 1.8 & 28 \\
  HConv~\cite{DBLP:conf/interspeech/ShihH24}
    & 2.1 & 4.3 & 7.7 & 1.4 & 2.9 & 11 \\
  Attentive fusion~\cite{DBLP:journals/corr/abs-2601-09322}
    & 1.9 & 4.0 & 5.8 & 1.6 & 2.7 & 24 \\
  \addlinespace[5pt]
  \multicolumn{7}{@{}l}{\textbf{Depth Heuristics}} \\[2pt]
  Middle layer (50\%)
    & 1.0 & 3.4 & 2.9 & 1.7 & 2.1 & 7 \\
  Final layer
    & 3.5 & 6.3 & 8.8 & 2.8 & 4.3 & 7 \\
  \bottomrule
  \end{tabular}
  \end{adjustbox}
  \caption{Mean gap (pp\,$\downarrow$) relative to the per-task oracle (best-performing single layer). \%\,${\geq}$\,Orac.: percentage of model--task pairs matching or exceeding it.}
  \label{tab:ls_comparison}
\end{table}

\vspace{2pt}

\noindent\textbf{Proxy ranking identifies strong candidate layers.} 
Evaluating only the top-ranked layer yields a $1.0$\,pp mean gap to the per-task oracle and already outperforms, on average, every trainable fusion baseline, consistent with prior work showing that a well-chosen single layer can match or exceed learned cross-layer aggregation~\cite{DBLP:conf/icassp/PasadSL23,DBLP:journals/corr/abs-2505-16306,DBLP:journals/taslp/YangCHLLWSCTHFCLCHTLLMWL24}. 
Expanding to the top three proxy-ranked layers reduces the mean gap to $0.4$\,pp and matches the oracle on $58\%$ of model--task pairs, providing robustness to imperfect rankings; in practice, $k{=}3$ probe runs suffice for near-oracle performance across model families and tasks, using roughly $4$--$16{\times}$ fewer supervised probe evaluations than exhaustive layer-wise scans; because selected layers are typically intermediate, deeper layers can also be pruned at inference.
Proxy-based selection also outperforms fixed-depth heuristics (Figure~\ref{fig:metrics-tasks}b). The middle layer is a competitive baseline for masked models, especially on flatter-profile tasks such as tagging, but becomes unreliable when performance is concentrated in a narrow, model-specific depth band, particularly in autoregressive models; in both families, the final layer is consistently weaker. In contrastive encoders, proxy selection typically favors the final layer, where most tasks peak, except on tonal tasks, where the PTE proxy selects an earlier optimum. Notably, PTE remains within $0.8$\,pp of the key-estimation oracle across model families and within $1$\,pp on most chord-recognition metrics, whereas fixed-depth heuristics perform poorly.

\vspace{2pt}

\noindent\textbf{Proxy-ranked layer fusion performs best overall.}
Among fusion methods, averaging the top-3 proxy-ranked layers yields the strongest performance ($p<0.001$ against trainable baselines; FDR-corrected Wilcoxon), with a mean gap of $0.3$\,pp to the oracle and improvements over it in $41\%$ of model--task pairs, reaching up to $2$\,pp. The advantage is largest for autoregressive backbones, where proxy-guided averaging stays within $0.1$\,pp of the oracle, with gains up to $1.5$\,pp, whereas every trainable fusion baseline has a gap of at least $2.8$\,pp. The same pattern holds for the similarity task, where no probe is trained and top-$k$ averaging outperforms all-layer averaging by $4.8$\,pp, suggesting that restricting fusion to proxy-identified layers is more effective than aggregating all layers indiscriminately for most tasks.
Learned fusion is also less robust in low-resource settings. For tasks with fewer than 1,000 training clips, every trainable baseline shows a mean gap of at least $5.6$\,pp, and on GTZAN-Genre (442 training clips), every trainable fusion method falls at least $10$\,pp below the oracle on average across models, suggesting that fusion weights are difficult to estimate reliably when probe supervision is limited.
Within the semantic family, the preferred strategy varies by task. Proxy-guided averaging performs best on multi-label tagging, consistently surpassing the best single-layer performance, whereas proxy-based layer selection is stronger on genre classification, emotion recognition, and similarity, where performance is concentrated around a single dominant layer.
Beat tracking and chord recognition are the only tasks where trainable multi-layer fusion consistently exceeds the best single layer: HConv improves by $2.7$--$4.4$\,pp and attentive fusion by $1.7$--$3.3$\,pp on average, with individual-model gains up to ${\sim}10$\,pp, suggesting that task-relevant cues are more distributed across depth for these downstream tasks than for others.


\section{Conclusions}\label{sec:conclusions}

We presented a systematic layer-wise analysis of intrinsic representation properties in music foundation models across masked, autoregressive, and contrastive pre-training paradigms. Our results show that no single property provides a universal proxy for layer quality across MIR tasks. Intrinsic dimension, curvature, and augmentation-conditioned metrics track layer-wise downstream transfer with task- and paradigm-dependent strength, but remain weak for tonal tasks, where pitch-transposition equivariance provides a more consistent proxy for tonal quality across model families.
Beyond analysis, these findings have practical value: intrinsic metrics can shortlist strong candidate layers, recovering near-oracle performance after only a few probe evaluations while avoiding exhaustive supervised layer-wise search. Notably, even a single proxy-ranked layer outperforms trainable multi-layer fusion methods on average, whereas fusion restricted to proxy-identified layers often yields further gains.

Our analysis is correlational: these metrics identify representation properties associated with stronger transfer, but do not establish whether directly manipulating them would affect downstream performance. Future work should therefore test whether explicitly regularizing such properties during pre-training can causally improve representation quality and transfer.
Likewise, because objective, supervision, data, architecture, and scale co-vary across existing models, disentangling their contributions to the observed properties will require controlled pre-training experiments.





\section{Acknowledgments}\label{sec:acknowledgments}

We thank Pedro Ramoneda and the anonymous reviewers for their constructive feedback.
This work is supported by the ``Cátedra IA y Música'' project (TSI-100929-2023-1), funded by the Secretaría de Estado de Digitalización e Inteligencia Artificial, the European Union-Next Generation EU funds, and BMAT Music Innovators; and by the ``IMPA'' project (PID2023-152250OB-I00), funded by MCIU/AEI/10.13039/501100011033/FEDER, UE.
We thankfully acknowledge the computer resources at MareNostrum and the technical support provided by Barcelona Supercomputing Center (IM-2024-2-0034).


\bibliography{ISMIRreferences}

%
%
%
%

\end{document}